\documentclass[fleqn,usenatbib]{mnras}

\usepackage{newtxtext,newtxmath}

\usepackage[T1]{fontenc}
\usepackage{ae,aecompl}

\usepackage{graphicx}	

\usepackage{epsfig}
\usepackage{natbib}
\usepackage{times}
\usepackage{threeparttable} 
\usepackage{booktabs}
\usepackage[euler]{textgreek}

\usepackage{xcolor}

\usepackage{epstopdf}

\newcommand{\chan}{\textit{Chandra}}
\newcommand{\swift}{\textit{Swift}}
\newcommand{\hubble}{\textit{Hubble Space Telescope}}

\newcommand{\xmm}{\textit{XMM-Newton}}

\newcommand{\hst}{\textit{HST}}

\newcommand{\Msun}{\mathrm{M}_{\odot}}
\newcommand{\lum}{\mathrm{erg~s}^{-1}}
\newcommand{\flux}{\mathrm{erg~cm}^{-2}~\mathrm{s}^{-1}}

\newcommand{\cnts}{\mathrm{counts~s}^{-1}}

\newcommand{\nh}{\mathrm{cm}^{-2}}
\newcommand{\nhp}{$N_\mathrm{H}$\,}

\newcommand{\exo}{EXO 0748--676}

\def \atel {\textit{ATel}}

\title[UV and X-ray observations of EXO 0748$-$676]{UV and X-ray observations of the neutron star LMXB EXO~0748$-$676 in its quiescent state}

\author[Parikh et al.]{A.S.~Parikh,$^{1}$
N.~Degenaar,$^{1}$\thanks{E-mail: degenaar@uva.nl} 
J.V.~Hern\'andez~Santisteban,$^{1,2}$
R. Wijnands,$^{1}$
\newauthor
I. Psaradaki,$^{1,3}$
E. Costantini,$^{1,3}$
D. Modiano,$^{1}$ and
J.M. Miller $^{4}$
\\
$^{1}$Anton Pannekoek Institute for Astronomy, University of Amsterdam, Science Park 904, NL-1098 XH Amsterdam, the Netherlands\\
$^{2}$School of Physics and Astronomy, University of St. Andrews, North Haugh, St. Andrews, Fife, KY16 933, Scotland, UK\\
$^{3}$SRON Netherlands Institute for Space Research, Sorbonnelaan 2, 3584 CA Utrecht, The Netherlands\\
$^{4}$Department of Astronomy, University of Michigan, 1085 South University Avenue, Ann Arbor, MI  48109, USA
}

\date{Accepted 2020 November 26. Received 2020 November 26; in original form 2020 June 26}

\pubyear{2019}
\hypersetup{draft}

\begin{document}
\label{firstpage}
\pagerange{\pageref{firstpage}--\pageref{lastpage}}
\maketitle

\begin{abstract}
The accretion behaviour in low-mass X-ray binaries (LMXBs) at low luminosities, especially at $< 10^{34} \mathrm{erg \ s}^{-1}$, is not well known. This is an important regime to study to obtain a complete understanding of the accretion process in LMXBs, and to determine if systems that host neutron stars with accretion-heated crusts can be used probe the physics of dense matter  (which requires their quiescent thermal emission to be uncontaminated by residual accretion).  
Here we examine ultraviolet (UV) and X-ray data obtained when \exo, a crust-cooling source, was in quiescence. Our {\it Hubble Space Telescope} spectroscopy observations do not detect the far-UV continuum emission, but do reveal one strong emission line, \ion{C}{iv}. The line is relatively broad ($\gtrsim 3500$~km~s$^{-1}$), which could indicate that it results from an outflow such as a pulsar wind. By studying several epochs of X-ray and near-UV data obtained with {\it XMM-Newton}, we find no clear indication that the emission in the two wavebands is connected. Moreover, the luminosity ratio of $L_{\mathrm{X}}/L_{\mathrm{UV}} \gtrsim 100$ is much higher than that observed from neutron star LMXBs that exhibit low-level accretion in quiescence. Taken together, this suggests that the UV and X-ray emission of \exo\ may have different origins, and that thermal emission from crust-cooling of the neutron star, rather than ongoing low-level accretion, may be dominating the observed quiescent X-ray flux evolution of this LMXB. 
\end{abstract}




\section{Introduction}
Low-mass X-ray binaries (LMXBs) are systems wherein a compact object accretes matter from a companion star, typically via an accretion disc. Many LMXBs exhibit transient behaviour and for these sources the accretion of matter onto the compact object is expected to (almost) stop during quiescence \citep[e.g.,][]{lasota2001disc,hameury2020}. The outburst luminosities exhibited by transient LMXBs hosting neutron stars (NSs) are typically on the order of 10$^{35}$--10$^{38}$ erg s$^{-1}$ in the 0.5--10 keV energy range \citep[e.g.,][]{yan2015xray}. In transient systems these luminosities drop to levels of 10$^{30}$--10$^{34}$ erg s$^{-1}$ when the outbursts cease \citep[see e.g.,][for a review]{wijnands2017review}.

During the brightest phases of an outburst, typically at luminosities $\gtrsim$10$^{36}$~erg~s$^{-1}$, the accretion flow takes on the form of a geometrically thin, optically thick disc \citep[e.g.,][]{shakura1973black,lasota2001disc}. The disc may be accompanied by a corona, a Comptonising medium of hot electrons, likely located close to the central compact object \citep[e.g.,][]{done2007modelling,2010LNP...794...17G}. The properties of the accretion flow close to the compact object are thought to change significantly towards lower luminosities by transitioning into a geometrically thick, optically thin flow also referred to as a radiatively-inefficient accretion flow \citep[RIAF; e.g.,][]{dubus2001disc,1994ApJ...428L..13N}. However, the conditions at which the transition from a standard disc to a RIAF take place are not very well understood, and hardly constrained by observations.

Studies of quiescent accretion flows are further exacerbated, as the inherent low luminosities exhibited in this state make it difficult to obtain high-quality data. The mere presence and properties of a disc, corona and/or RIAF at these luminosities is not definitively known \citep[e.g.][]{hynes1999hubble,hynes2005observational,cackett2013_cenx4}, so it is not clear if, and how, accretion takes place.  Accretion onto the compact object could potentially come to a halt in quiescence, but there could also be ongoing inward motion of the material through a disc/corona/RIAF, perhaps still enabling low-level accretion onto the compact object. Studies in quiescence are important to understand accretion at low luminosities and contribute to our overall understanding of accretion processes in LMXBs.

Apart from studying accretion, transient NS LMXB systems can also be used as laboratories to probe the properties of dense matter present in NS crusts and cores \citep[e.g.,][]{brown1998crustal,cumming2017lower,wijnands2017review}. During outbursts, the accreted matter compresses the NS crust and induces reactions such as electron captures, neutron emissions, and density-driven fusion reactions \citep[e.g.,][]{haensel1990non,haensel2008models,steiner2012deep}. These reactions release heat in the crust, disrupting its thermal equilibrium with the core. Furthermore, observations of NS LMXBs have indicated that, in addition to these theoretically predicted heating reactions, heat may also be released at shallow depths in the crust by an unknown `shallow heating' process \citep[e.g.,][]{brown2009mapping,degenaar2011evidence,degenaar2014probing}. 

With the cessation of an outburst, i.e., active accretion, the compression-induced heating reactions occurring in the NS crust are expected to stop. As a result, in quiescence the crust cools to reinstate the thermal equilibrium with the core. Monitoring this cooling evolution and comparing it to theoretical models, has been used to investigate the behaviour of the dense matter present in NS crusts \citep[e.g.,][]{wijnands2001chandra,rutledge2001,shternin2007,brown2009mapping,page2013forecasting,turlione2015,ootes2018}. So far, such crust-cooling studies have been carried out for ten NSs in LMXBs by examining their X-ray spectral evolution \citep[see][for a review]{wijnands2017review}. The quiescent spectra of NSs exhibiting crust cooling are typically characterised by a prominent thermal emission component (often contributing $>$70~per cent of the total 0.5--10 keV flux) that is interpreted as heat radiation from the NS surface. 

For several NSs in LMXBs there is evidence that there is some form of low-level accretion occurring, either continuously or in spurts, at similarly low luminosities as are exhibited by crust-cooling sources in quiescence \citep[e.g., EXO 1745$-$248, Cen X-4, Aql X-1, XTE J1701--462;][]{fridriksson2011,degenaar2012strong, bernardini2013,cotizelati2014}. Notably, the spectra of these sources are characterised by a prominent power-law shaped emission component that, when detailed studies are possible, appears to extend to energies $>$10 keV \citep[][]{chakrabarty2014hard}. Often this power-law emission component occurs alongside a thermal emission component. Indeed, theoretical calculations suggest that low-level accretion onto a NS can produce a composite spectrum of thermal and power-law like emission components \citep[e.g.,][]{deufel2001}. 

Several crust-cooling sources also require a power-law shaped component, in addition to thermal surface emission, to adequately fit their quiescent spectra \citep[e.g.,][]{degenaar2009chandra,wijnands2015low,parikh2017potential}. This could potentially indicate that low-level accretion is ongoing in these systems. If true, it is not trivial to disentangle contributions of low-level accretion and crust-cooling emission to the thermal emission spectra. This then complicates our determination of the crust-cooling evolution and thereby the inferred NS physics. Constraining the properties of low-level accretion in LMXBs is therefore also important to correctly infer the physics of the NS crust and core. 

In addition to examining the X-ray spectra, accretion can also be probed at ultraviolet (UV) wavelengths. The surface temperatures of NSs in LMXBs are generally such ($\simeq10^6-10^7$~K) that their cooling emission is not expected to significantly contribute to their UV emission. Also, the inherent emission from the late-type companion star contributes negligibly to the UV emission of the system. However, the accretion stream, arising from the Roche lobe overflow of the donor star, is expected to be UV bright, hence making the UV a suitable wavelength regime to probe the quiescent accretion flows. Previous studies of quiescent LMXBs have indicated that the UV emission arises from small emitting radii with unexpected high temperatures, suggesting that this does not come from the bulk of the disc but rather from the impact point of matter from the companion star hitting the accretion disc \citep[e.g.,][]{froning2011}, or X-ray irradiation of the inner accretion disc or companion star \citep[e.g.,][]{hynes2012,cackett2013_cenx4,bernardini2013}. 

A further indicator of possible ongoing accretion towards the NS can be derived from the correlation exhibited between the X-ray and UV luminosities in quiescence \citep[e.g.,][]{hynes2012}. In particular, if the UV and X-ray luminosities are similarly bright, the UV emission is too high to only arise from irradiation of the disc (if present) and/or the companion star by X-rays from the central compact object (or from a hot inner accretion disc, if present).

Most NS LMXBs cannot be studied in the UV as they exhibit strong interstellar extinction along the line of sight. However, a few LMXBs exhibit less strong extinction, allowing us to probe for signatures of low-level accretion in the UV. One such source is the crust-cooling source EXO 0748$-$676.

\subsection{The NS LMXB \exo}
\label{sec_intro_exo}
The NS LMXB \exo\ (UY Vol) was discovered almost three decades ago \citep[][]{parmar1985}. It exhibited X-ray eclipses (lasting $\simeq$500~s), indicating that the binary is viewed at high inclination ($i\simeq75^{\circ}-83^{\circ}$). Its orbital period was determined to be $\simeq$3.8~hr \citep[][]{parmar1986,wolff2008c}. The source displayed thermonuclear X-ray bursts, which indicated the NS nature of the primary and resulted in a distance estimate of $\simeq$7.4 kpc \citep[][]{wolff2005}.

\exo\ exhibited an accretion outburst that is estimated to have started between 1980 and 1984 \citep{degenaar2009chandra}. It remained active until September 2008 \citep[][]{wolff2008,wolff2008b,hynes2008,torres2008}. Observations of the source in quiescence indicate that the crust of the NS was heated up during outburst and exhibited cooling in quiescence \citep[e.g.,][]{degenaar2009chandra,degenaar2011further,degenaar2014probing,diaztrigo2011,cheng2017cooling}. However, more recent quiescent observations, obtained $>$2000~d after the end of its accretion outburst, indicate a rise in the observed effective NS temperature \citep{parikh2020dip}. This rise is highly surprising for a crust-cooling source as it is expected to exhibit a continuous decay in the temperature rather than a rise. In this work, we investigate if low-level accretion occurs in the quiescent state of \exo.

\section{Observations and data analysis}
We examine UV spectroscopic and imaging data of \exo. We also compare these UV data with (quasi-)simultaneous X-ray data. De-reddened photometric fluxes and magnitudes were calculated by assuming $E(B-V)=0.06$ \citep{hynes2006} and the Galactic extinction curves of \citet{cardelli1989}. The de-reddening has been carried out in \textsc{Python} using the \textsc{extinction} package.\footnote{\url{https://extinction.readthedocs.io/en/latest/}} All uncertainties reflect 1$\sigma$ confidence levels.

\subsection{\hubble\ observations}
\label{subsec:hst}

\subsubsection{Far-UV spectroscopy}
\label{sect_fuv_method}
We obtained far-UV (FUV) spectroscopic observations of \exo\ in quiescence using the  Cosmic Origins Spectrograph \citep[COS;][]{green2012} on board the {\it Hubble Space Telescope} ({\it HST}) under the GO program 13108 (PI: Degenaar). The source was targeted for 9 orbits divided over 2 visits, starting on 2013 March 28 (MJD 56379). The total on-source exposure time amounted to $\simeq$7.4~hr. The COS was operated using the low-resolution G140L grating with segment B switched off. This provided FUV coverage in the wavelength range of $\simeq$1118 -- 2251~\AA\ with a resolving power of $\uplambda /\Delta \uplambda \simeq 2500-3500$~\AA. For each exposure, the grating was stepped through different fp-pos positions to minimize the fixed pattern noise. All the data were taken in TIME-TAG mode. These COS data were retrieved from the Multi-Mission Archive at STScI\footnote{\url{https://archive.stsci.edu/hst/search.php}}. The downloaded data have been pre-processed with \textsc{calcos} (version 3.3.7).
  
To contextualise the quiescent FUV spectra of \exo, we compared it to the FUV spectrum obtained during outburst. This outburst spectrum was obtained in 2003 February 18 (MJD 52688) using the Space Telescope Imaging Spectrograph \citep[STIS; e.g.,][]{2015IAUGA..2255542E}. The source was observed for $\simeq$9~hr over 2 complete orbits in the continuous viewing zone. The observations were carried out using the FUV MAMA detector \citep[][]{charles1995performance} and the G140L grating. The data were taken in the TIME-TAG mode. The STIS data were retrieved from the Multi-Mission Archive at STScI and were pre-processed with \textsc{calstis} (version 3.4.2). Further details of these data are provided in \citet{hynes2006} and \citet{pearson2006}.

Spectral analysis on these FUV data was performed on the products downloaded from the archive. The various one-dimensional spectra, corresponding to each quiescent epoch, were combined using the \textsc{splice} task in \textsc{IRAF}\footnote{\url{https://iraf-community.github.io/install.html}} \citep[e.g.,][]{doug1986iraf}. These spectra were further analysed using the \textsc{specutils} package\footnote{\url{https://specutils.readthedocs.io/en/stable/index.html}}  \citep[][]{2019specutils} in \textsc{Python} and are shown in Figure \ref{fig:spec_exo}. Line widths were calculated based on Gaussian fits using \textsc{line{\_}flux}, part of the \textsc{specutils} package in \textsc{Python}.

\subsubsection{Near-UV photometry}
\label{sect_nuv_phot_method}
Each of the two \hst\ visits of \exo\ during quiescence were preceded by acquisition imaging observations of $51$~s each, using mirror A and the PSA aperture. This provided sensitivity in the UV range of $\simeq$1600 -- 3300~\AA, peaking at $\simeq$2300~\AA. 

We estimated a Near-UV (NUV)\footnote{In this paper, we define NUV to encompass all data in the wavelength range 1600 -- 3000~\AA. These include the NUV data discussed in Sections \ref{sect_nuv_phot_method}, \ref{subsec:swift}, and \ref{subsec:xmm}.} magnitude for \exo\ from these acquisition images. Source events were extracted from an aperture of radius $0.10''$ using \textsc{Python}. Furthermore, an aperture of three times that size was used to estimate the background. The detected count rates were converted to fluxes using the count rate to flux conversion factor appropriate for our instrumental set-up. This conversion factor, reported in the headers of the acquisition images, was found to be $1~\cnts =4.82\times10^{-18}~\flux$~\AA$^{-1}$.

\subsection{\swift\ observations}\label{subsec:swift}
\exo\ was observed with the X-ray telescope \citep[XRT;][]{burrows05} on board the {\it Neil Gehrels Swift Observatory} \citep{gehrels2004swift}. A $\simeq$7.8~ks observation was carried out on 2013 March 23 (MJD 56374), i.e., $\sim$5~d before the \hst\ UV observations were obtained. We reduced the \swift\ data\footnote{Obtained from the {\it Swift} archive, found at \url{https://heasarc.gsfc.nasa.gov/cgi-bin/W3Browse/swift.pl}} using the \textsc{Heasoft} package (version 6.13), by re-processing the data using the \textsc{xrtpipeline}. We extracted the spectrum of the source with \textsc{XSelect} using a circular region with a radius of $35''$ centred on \exo. A surrounding annular region, having an inner and outer radius of $100''$ and $200''$ respectively, was used to obtain a background spectrum. The spectral data were grouped into bins with a minimum of 10 photons per bin using the \textsc{Heasoft} tool \textsc{grppha}. The spectrum was fit using \textsc{XSpec}.

Since \exo\ hosts a cooling NS crust \citep[e.g.,][see the references therein for more details]{parikh2020dip}, we fitted the XRT spectrum with \textsc{nsatmos} \citep[a NS atmosphere model;][]{heinke2006hydrogen}. The source distance was set to $D=7.4$~kpc \citep[e.g.,][]{galloway2008} and the NS mass and radius to $M=1.6~\Msun$ and $R=12$~km, respectively. Given the limited statistics of the XRT data, we fixed several of the \textsc{nsatmos} parameters to the values obtained from fitting high-quality quiescent \chan\ and \xmm\ spectra of \exo\ \citep[see][for details, where the same parameters have been fixed]{parikh2020dip}. An additional power-law component was needed to fit some high-quality quiescent spectra of this source \citep[][]{parikh2020dip}. Thus, we also use this component for the XRT spectral fitting here. The power-law index was fixed to $\Gamma=1.0$ and the normalisation was free to vary. In addition, we fix the column density to \nhp = $4.3 \times 10^{20}\ \nh$, as determined from fitting high quality data, presented in \citet{parikh2020dip}.

Alongside the XRT, {\it Swift} hosts the Ultraviolet Optical Telescope \citep[UVOT;][]{roming2005swift}, with NUV photometric and spectroscopic capabilities. We examined all archival UVOT data. We processed the data using the \textsc{uvousource} tool. However, the source was not detected by the UVOT during any quiescent observations. Furthermore, since the upper limits obtained from these data were not constraining we do not discuss them further.

\subsection{\textit{XMM-Newton} observations}\label{subsec:xmm}
The {\it X-ray Multi-Mirror Mission} or {\it XMM-Newton} hosts instruments with observation capabilities in the X-ray and NUV wavelength regimes. This allows for correlated X-ray--UV studies as instruments observing in both wavelength ranges operate simultaneously. So far, \exo\ has been observed 6 times during quiescence using {\it XMM-Newton}. These observations were downloaded from the {\it XMM-Newton} archive.\footnote{\url{http://nxsa.esac.esa.int/nxsa-web/\#home}}

The X-ray data, obtained using the European Photon Imaging Camera \citep{struder2001european,turner2001european}, were processed by extracting the spectra and fitting them collectively in \textsc{XSpec}. The details of the spectral extraction and spectral fitting are provided in Section~2 of \citet{parikh2020dip}. 

The NUV photometric data, obtained using the Optical Monitor \citep[OM;][]{mason2001OM}, were only available for five of the six observations. This is because the data for observation 3 (corresponding to observation Id 0605560501) were not obtained due to a technical error \citep{diaztrigo2011}. For the available data, the NUV fluxes were extracted using the OM tasks \textsc{omichain} and \textsc{omfchain} in \textsc{sas}, as was appropriate when the photometric data were obtained in the imaging mode and in the fast mode, respectively. We examined the fluxes corresponding to the full exposure (for each available filter, from the output files produced by \textsc{omichain} and \textsc{omfchain}) from each entire observation, in order to use the most constraining flux estimate for a given epoch. Furthermore, this is because we do not expect to be able to detect any orbital variations in the flux due to the low UV luminosities exhibited by \exo\ \citep[as was shown by][during the earliest quiescent UV observations obtained]{diaztrigo2011}. \exo\ was not detected during several OM observations (see Table \ref{tab_xray_uv}). In case of non-detections, we used the faintest detected source in the associated observation as an upper limit estimate for \exo\ (see Table \ref{tab_xray_uv}). This was done by examining the output files produced by \textsc{omichain} and \textsc{omfchain}. Since the default detection threshold is 3$\sigma$, our obtained upper limits correspond to this significance level.

\section{Results}

\begin{figure*}
	\includegraphics[width=\textwidth]{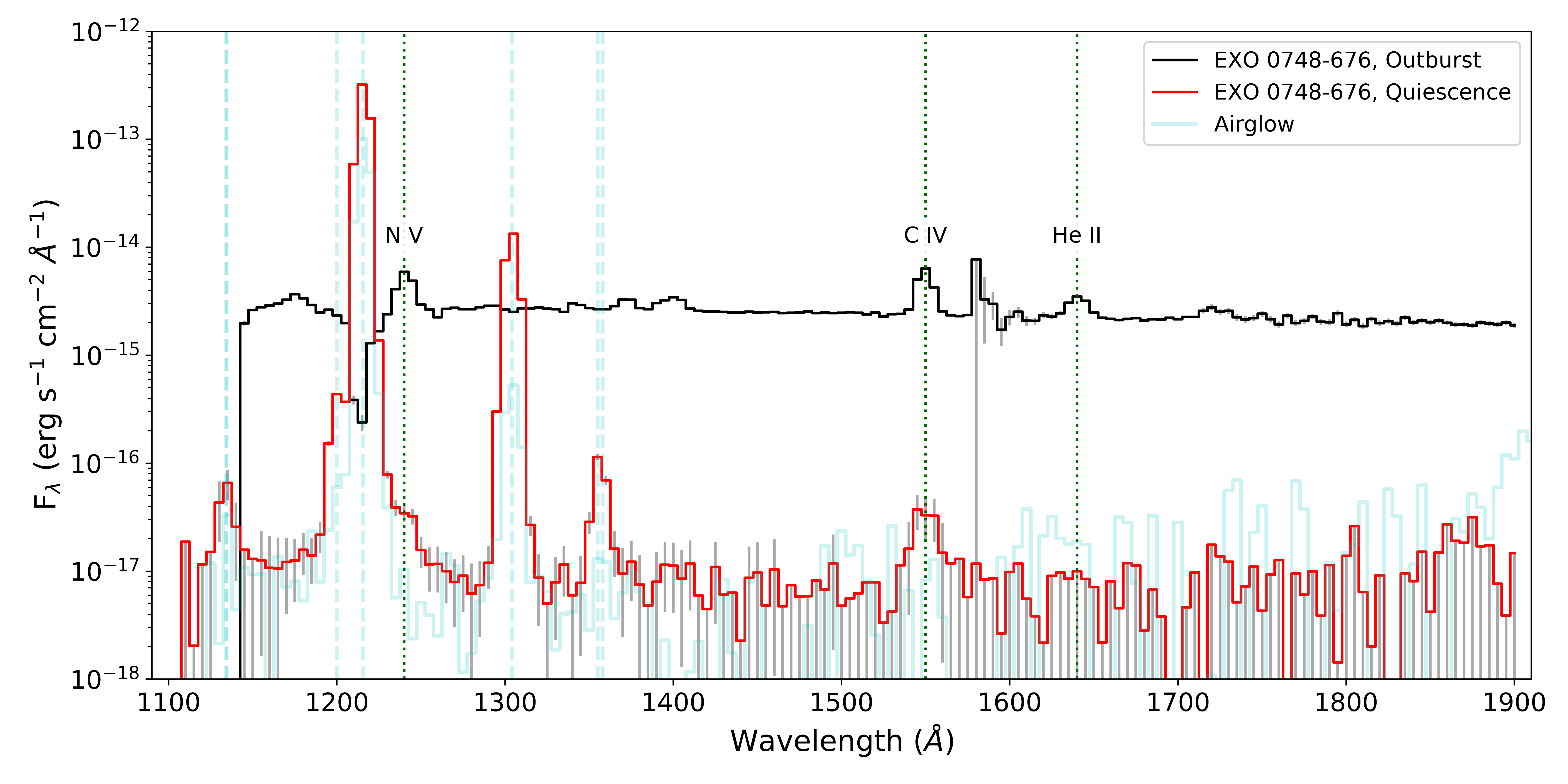}
    \caption{Co-added FUV spectra of \exo\ obtained using \hst\ during two epochs --- during outburst (using STIS, black) and during quiescence (using COS, red). The respective error bars are shown in grey. Both spectra have been re-binned to the same spectral resolution (of $\Delta\uplambda$ = 5 \AA). The vertical, dotted, green lines indicate the strong emission lines seen in the outburst spectra, as identified by \citet{pearson2006}. The spectrum shown in blue is an example of the airglow spectrum as observed using the COS instrument. The vertical, dashed, blue lines are often-observed airglow emission lines exhibited during observations obtained using COS \citep[e.g.,][]{froning2014}. The STIS observation does not suffer from contamination due to airglow. }
    \label{fig:spec_exo}
\end{figure*}

\subsection{The \hst\ FUV spectra}

\subsubsection{Quiescent spectrum}
Despite the low extinction towards \exo, the quality of the quiescent FUV spectra obtained by \hst\ during individual orbits was low. This was due to the inherently low source flux in quiescence. The signal-to-noise ratio was inadequate to perform any phase-resolved spectral analysis which is necessary to investigate signatures of the orbital motion. We, therefore, combined individual exposures into a uniform wavelength grid (with $\Delta\uplambda$ = 5~\AA) and applied an error-weighted average using the \textsc{spectres} package \citep{2017arXiv170505165C} in \textsc{Python}. The resulting averaged spectrum is shown in red in Figure~\ref{fig:spec_exo}. Although we combined the different COS exposures, the resulting spectrum was still of low signal-to-noise and there was no continuum emission detected above $\simeq10^{-17}$ erg cm$^{-2}$ s$^{-1}$ \AA$^{-1}$, preventing us from obtaining any continuum flux measurement from the FUV spectral data.

\subsubsection{Outburst spectrum}
We aim to qualitatively compare the outburst spectrum to that obtained during quiescence and are not interested in probing the orbital variations observed during the outburst. Thus, the individual STIS spectra were combined into one spectrum and as a result, this blurs out the orbital variations observed in outburst \citep[][]{pearson2006}. This has been done as described in Section \ref{sect_fuv_method}. To ensure direct comparison with the quiescent spectrum, the combined outburst spectrum was re-binned to the same resolution (of $\Delta\uplambda$ = 5~\AA) as that of the combined quiescent spectrum. This is shown in Figure~\ref{fig:spec_exo}, in black. The dotted, green, vertical lines indicate the significant emission lines in the outburst spectrum, as identified by \citet{pearson2006}. The emission observed near 1580~\AA\ is an artefact and does not correspond to any line \citep[][]{pearson2006}.

\subsubsection{Airglow observed by the COS instrument}
Airglow is the contribution of emission from the Earth's atmosphere. The quiescent FUV spectrum is shown alongside an example of a spectrum obtained using the same COS instrument when the terrestrial airglow was strongly observed.\footnote{The different observations of airglow by the \hst/COS can be found at \url{https://www.stsci.edu/hst/instrumentation/cos/calibration/airglow}. We show the airglow spectrum corresponding to the dataset LBW3E4060, observed on 2012 August 10 (MJD 56149).} This is shown in blue in Figure~\ref{fig:spec_exo} and has been re-binned to have the same spectral resolution as the quiescent source spectrum. As can be seen, this airglow spectrum roughly matches the strong emission lines seen around 1200, 1215, and 1304~\AA\ in our quiescent spectrum of \exo. Evidence from other airglow observations, obtained using COS, show that the emission line observed around 1356~\AA\ is also a result of airglow \citep[e.g.,][]{froning2014}. The prominent wavelengths at which airglow emission lines are expected is shown using vertical, dashed, blue lines. The outburst spectra were taken with a different instrument and do not suffer from airglow \citep[][]{pearson2006}.

Discarding any emission lines due to airglow, we find that our quiescent observation of \exo\ only exhibits one significant emission line around 1550~\AA, identified to be the \ion{C}{iv} line by \citet{pearson2006}. Due to the lack of a continuum detection only the lower limit of the \ion{C}{iv} line could be constrained during quiescence. This \ion{C}{iv} emission line has a width of $\simeq$19.5 and $\gtrsim$18.2~\AA\ in the spectrum obtained during outburst and quiescence, respectively.

We also investigated if any other lines that were significant during the outburst were present in quiescence. The red wing of the emission line in the quiescent spectra, observed around 1220~\AA, exhibits an excess as compared to the corresponding airglow emission line. This may be indicative of a \ion{N}{v} emission line from the source, as was seen during outburst \citep[see Figure~\ref{fig:spec_exo}; see also][]{pearson2006}. This excess is relative to our choice of the example airglow emission spectrum shown. However, it is not trivial to constrain any excess in the spectrum due to emission by the \ion{N}{v} line in quiescence.

\subsection{The ratio of the X-ray and UV luminosities}
\label{subsec:uv}
We examined quasi-simultaneous epochs of X-ray and NUV data to investigate the correlation between emission in these two energy bands.

\subsubsection{{\it HST} NUV photometry}\label{sect_res_hst}
Visual inspection of the NUV acquisition images obtained by the \hst/COS show that \exo\ was faintly detectable. Examining these images, we estimated an average (background-subtracted) source count rate of $\simeq 0.16~\cnts$. Using the appropriate count rate to flux conversion factor (see Section \ref{sect_nuv_phot_method}), we estimated a flux density of $F_{\uplambda} \simeq 7.3\times10^{-19}~\flux$~\AA$^{-1}$, and an AB magnitude of $\simeq 26.1$~mag. The de-reddened flux density and magnitude were found to be $F_{\uplambda}\simeq 1.2\times10^{-18}~\flux$~\AA$^{-1}$ and $\simeq 25.6$~mag, respectively. 

By multiplying the flux with the central wavelength of the instrument passband \citep[$2300$~\AA;][]{2012cosi.book}, we estimate a de-reddened UV flux of $F_{\mathrm{UV}}\simeq 2.8 \times10^{-15}~\flux$. For a distance of 7.4~kpc, this translates into a luminosity of $L_{\mathrm{UV}}\simeq 1.8\times10^{31}~\lum$. We note, however, that this approach effectively assumes that the UV spectrum is flat over the instrument passband, which is not necessarily the case. We investigate the effect of this assumption in Section~\ref{sec:UV_bb}.

\subsubsection{{\it Swift} X-ray data}
\label{sect_res_swift}
The model described in Section~\ref{subsec:swift} provided a good description of the X-ray spectral data ($\chi_{\nu}^2=0.97$ for 31 d.o.f.). We obtained $kT^{\infty}_\mathrm{eff}=111.3 \pm 2.0 $~eV and the resulting unabsorbed flux was $F_{\mathrm{X}} = (5.3 \pm 0.4) \times10^{-13}~\flux$, in the 0.5--10~keV  energy range. At a distance of 7.4 kpc, this corresponds to a luminosity of $L_{\mathrm{X}} =  (3.5 \pm 0.3) \times10^{33}~\lum$. Thus, using our quiescent quasi-simultaneous {\it Swift} X-ray and \hst\ NUV observations, we obtained a luminosity ratio of $L_{\mathrm{X}}/L_{\mathrm{UV}} \simeq 200$. 

\subsubsection{{\it XMM-Newton} X-ray and UV data}\label{sect_res_xmm}
\exo\ was detected two times in quiescence using the OM with the U filter (see Table~\ref{tab_xray_uv}). Upper limits were obtained during three other epochs --- once using the UVW1 filter and twice using the U filter. The results of the OM photometry is listed in Table~\ref{tab_xray_uv}. The source was detected in X-ray during all the {\it XMM-Newton} epochs reported here (i.e., those with accompanying OM observations, see Section \ref{subsec:xmm}). Details about the X-ray luminosities are presented in \citet[][see their table~1]{parikh2020dip}. 

If we calculate the UV luminosity by multiplying the inferred UV fluxes with the filter width (but see Section~\ref{sec:UV_bb}), we obtain $L_{\mathrm{X}}/L_{\mathrm{UV}}$ ratios of $\simeq$100 from the {\it XMM-Newton} data (see Table~\ref{tab_xray_uv}). This is on the same order of magnitude as the ratio inferred for the epoch of quasi-simultaneous {\it HST} and {\it Swift} observations (see Section~\ref{sect_res_swift}).

\subsubsection{The effect of the unknown UV spectral shape}
\label{sec:UV_bb}
In Section~\ref{sect_res_hst} and~\ref{sect_res_xmm}, we estimated $\lambda L_{\lambda}$ for all UV observations in order to compare the X-ray and UV luminosity ratio of \exo\ with that of the eight LMXBs studied in \citet{hynes2012}. However, the UV spectrum is likely not flat and may significantly change across the wavelength range that we study. To probe the magnitude of this effect, we folded a stellar atmosphere spectra with appropriate effective temperatures \citep[][]{allard2012}\footnote{Available at https://phoenix.ens-lyon.fr/Grids/BT-Settl/} through the relevant UV filter and then determined the expected ratio of fluxes in different wavebands for each observation.\footnote{The response for the \hst/COS acquisition image is not available, so we took the WFC3/UVIS F218W filter for this as a proxy.} For this exercise we used two different temperatures, of 5\,000~K and 13\,000~K, based on the UV-optical spectral energy distributions (SEDs) of the eight quiescent LMXBs studied in \citet{hynes2012}. We scaled the two different SEDs by performing synthetic photometry on them and comparing them to the unabsorbed fluxes from Table~\ref{tab_xray_uv}. We then integrated the fluxes and compared these to a flat spectrum (i.e., the implicit assumption when calculating $\lambda L_{\lambda}$). 

We show the results of these calculations for each observation in Figure~\ref{fig:spec_effect}. The SED shape has the largest effect on the UV luminosity calculation for the bluest filter. For the \hst\ acquisition image, the ratio of the integrated flux from the reddest SED (i.e., the one with an effective temperature of 5\,000~K) compared to a flat spectrum is a factor $\simeq$1.5. The magnitude of the effect in the U band is much smaller, a few per cent at most (see Figure~\ref{fig:spec_effect}).

\subsubsection{$L_{\mathrm{X}}$ versus $L_{\mathrm{UV}}$}
\begin{figure}
 \begin{center}     
  \includegraphics[width=\columnwidth]{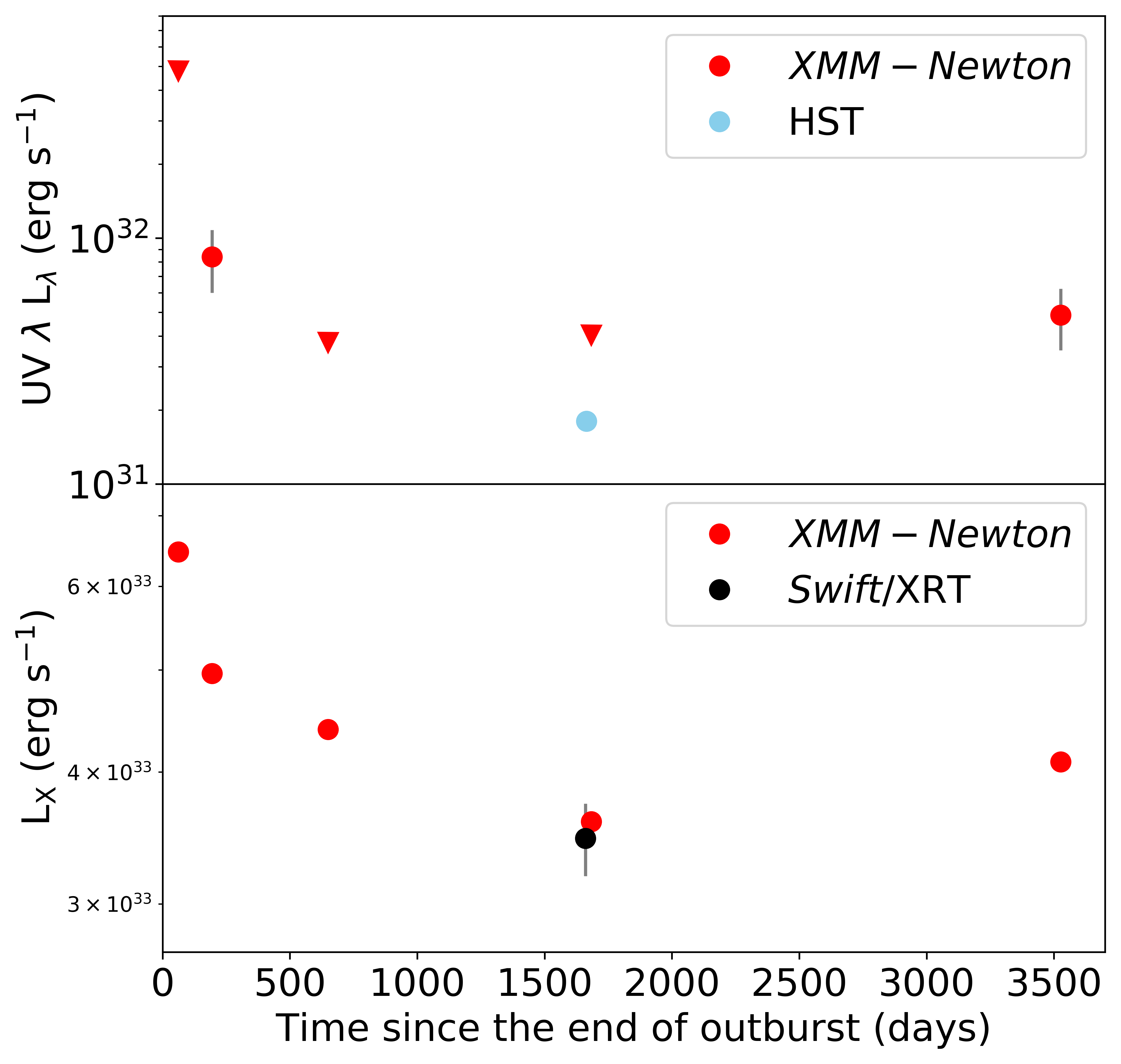}
    \end{center}
\caption{The top and bottom panels show the quasi-simultaneous evolution of the UV and X-ray luminosity of \exo\ during quiescence. The data from the quasi-simultaneous \hst\ and {\it Swift} observations are shown in blue and black, respectively. The red data indicate those observations obtained using {\it XMM-Newton}. All upper limits are marked by downward-facing triangles. We note that the plotted UV data do not take into account uncertainties on the UV spectral shape, which most strongly affects the \hst\ data point (see Section~\ref{fig:spec_effect}); caution should therefore be taken in comparing this point to the \xmm\ data (the first of which was taken with the UVW1 filter, the others with the U filter).} 
 \label{fig:uvxt}
\end{figure}

\begin{figure}
 \begin{center}     
  \includegraphics[width=\columnwidth]{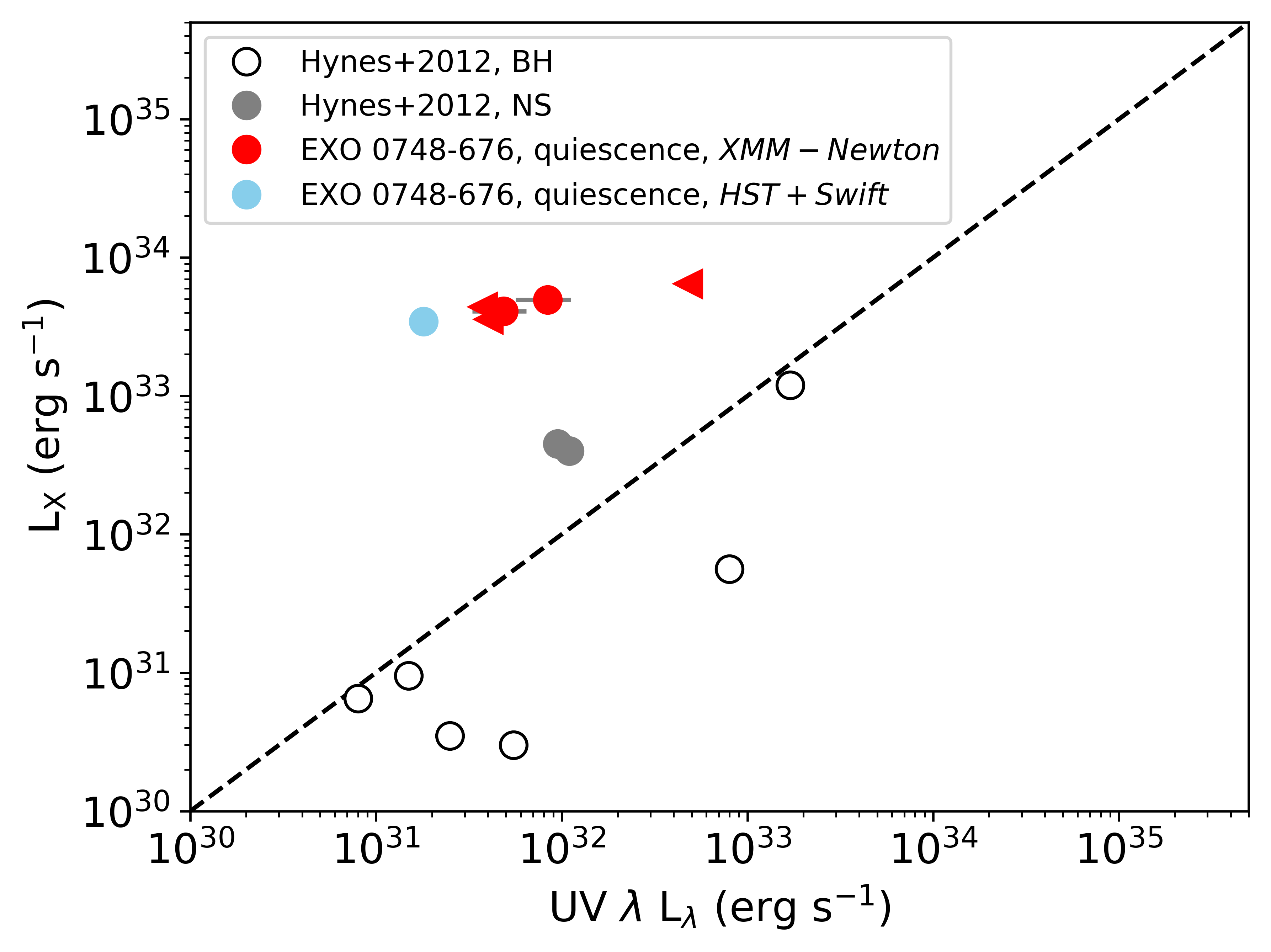}
    \end{center}
\caption{ The observed X-ray versus UV luminosity of \exo\ during six quiescent epochs is shown. Five epochs were obtained using {\it XMM-Newton} (shown in red). One epoch was observed using \hst\ and the {\it Swift}, and is shown in blue. 
We note that these data points do not take into account the unknown UV spectral shape (see Section~\ref{fig:spec_effect}). 
We compare \exo\ to eight other quiescent LMXBs studied in \citet{hynes2012}. The six systems hosting BHs are shown by open black circles and represent GRO~J0422$+$32 (V518 Per), MM Vel, XTE J118$-$40, A0620$-$00, GU Mus, and V404 Cyg (listed in ascending order of the UV luminosity). The two NS systems are shown by filled grey circles and correspond to Aql X-1 and Cen X-4. All upper limits are marked by triangles facing the relevant direction. }
 \label{fig:uvx}
\end{figure}

The unabsorbed X-ray luminosities in the  0.5--10 keV energy range, along with the (quasi-)simultaneously obtained UV luminosities (corrected for extinction) are summarised in Table~\ref{tab_xray_uv} and shown in Figures~\ref{fig:uvxt} and~\ref{fig:uvx}. In the X-rays, \exo\ shows a steady decrease by about 40~per cent over the $\simeq$9.5~yr time span covering the observations \citep[see also][]{parikh2020dip}. In the UV band, such long-term variations are not clearly seen.

\exo\ was not detected in the UV during the first \xmm\ epoch, when the OM was used with the UVW1 filter. However, the obtained upper limit is not constraining and much higher than the U-band luminosities measured at later times. During the second and last \xmm\ epochs, \exo\ is detected in the U filter at luminosities of $\simeq8.5$ and $5\times 10^{31}~\lum$, respectively, but with large errors; the two data points are consistent within $2~\sigma$ even without taking into account the additional uncertainty of the UV spectral shape in this band (see Section~\ref{sec:UV_bb}). The X-ray luminosity corresponding to these two data points differs by about 20~per cent. \exo\ is not detected in the U-band in the other two observations, but the obtained upper limits are not significantly smaller than the U-band luminosities inferred for the two detections. While the inferred luminosity from the {\it HST} acquisition image is a factor of a few lower than the U-band luminosities measured with \xmm\ (see Table~\ref{tab_xray_uv} and Figure~\ref{fig:uvxt}), the unknown UV spectral shape introduces significant errors in this band (see Section~\ref{sec:UV_bb}). The available data thus does not indicate that there are strong variations in the UV emission of \exo\ over the $\simeq$9.5-yr observing period. However, the uncertainties are large and variations of similar magnitude as seen in the X-rays ($\simeq$40~per cent) cannot be excluded. 

Figure \ref{fig:uvx} also shows the X-ray versus UV luminosity for other NS and BH LMXBs observed in quiescence, as reported by \citet{hynes2012}. It is clear from this plot that while the UV luminosity of \exo\ is in the same realm as that observed for other LMXBs, its X-ray luminosity, and hence its X-ray/UV luminosity ratio, is much larger than that of any of the other systems.

\begin{table*}
\caption{Log of the (quasi-)simultaneous  UV and X-ray observations of \exo\, in quiescence.}
\label{tab_xray_uv}
\begin{tabular}{lllcclcl}
\hline
Epoch & MJD & Observatory & UV band & Central UV & UV $\uplambda\mathrm{L}_\uplambda$ & Unabsorbed L$_\mathrm{X}$ & L$_\mathrm{X}$/L$_\mathrm{UV}$ \tabularnewline
 &  &  & & wavelength$^{a}$ (\AA) & ($\times 10^{31}$ erg s$^{-1}$) & ($\times 10^{32}$ erg s$^{-1}$) & ratio \tabularnewline
 \hline
1 & 54776.4 & {\it XMM-Newton} & UVW1 & 2910 & $<$47.8 & 64.7$\pm$0.5 & $>$13.5 \tabularnewline
2 & 54908.0 & {\it XMM-Newton} &U & 3400 & 8.4$\pm$2.4 & 49.6$\pm$0.3 & $\simeq$59.1 \tabularnewline
3 & 55364.2 & {\it XMM-Newton} &U & 3400 & $<$3.8 & 43.9$\pm$0.3 & $>$115.5 \tabularnewline
4 & 55376.5$^{b}$ & \hst\ \& {\it Swift} & NUV & 2300 & $\simeq$1.8 & 34.7$\pm$2.6 & $\simeq$243.9 \tabularnewline
5 & 56397.2 & {\it XMM-Newton} &U & 3400 & $<$4.0 & 35.9$\pm$0.2 & $>$89.8 \tabularnewline
6 & 58240.3 & {\it XMM-Newton} &U & 3400 & 4.9$\pm$1.4 & 40.9$\pm$0.2 & $\simeq$83.5 \tabularnewline
\hline
\multicolumn{8}{p{13cm}}{We have assumed $D=7.4$~kpc and $E(B - V) = 0.06$. We caution that the UV luminosities quoted here assume a flat spectral shape; see Section~\ref{sec:UV_bb} for a discussion on the impact of the unknown spectral shape.}\tabularnewline
\multicolumn{8}{p{13cm}}{$^{a}$Obtained from \url{https://www.swift.ac.uk/analysis/uvot/filters.php}.} \tabularnewline
\multicolumn{8}{p{13cm}}{$^{b}$The MJD for this epoch has been determined by averaging the MJD of the \hst\ and {\it Swift} observations} \tabularnewline
\end{tabular}
\end{table*}

\section{Discussion}
Investigating quiescent spectra of LMXBs allows us to understand the accretion behaviour in these systems at low luminosities.  In addition, quiescent studies of accretion-heated NS crusts in LMXBs provide us with the opportunity to probe the properties of dense matter physics. We have examined the crust-cooling source \exo\ in quiescence. Early in quiescence, this source has behaved like a typical crust-cooling source. However, recently obtained observations indicate an unexpected temperature rise (see Section~\ref{sec_intro_exo}). This rise in temperature may be a result of increased low-level accretion in quiescence, although studies of the X-ray spectra indicate that this is likely not the case \citep[see][for more details]{parikh2020dip}. Here we have further investigated the possibility of ongoing accretion in quiescence in \exo\ by examining the quiescent FUV spectrum and several epochs of (quasi-)simultaneous X-ray and NUV data.

\subsection{ The quiescent FUV spectrum}
Due to the inherently low quiescent FUV flux of EXO 0748--676, all FUV spectroscopic data had to be stacked to increase the signal to noise and hence no phase-resolved study could be performed. We do not detect FUV continuum emission from the source, but there is one prominent emission line detected, \ion{C}{iv} ($\simeq$1550~\AA), for which we measure a width of $\gtrsim 18.2$~\AA\ ($\gtrsim$3523~km~s$^{-1}$).

The FUV spectrum of this source obtained during the outburst also shows a prominent \ion{C}{iv} line (alongside other emission lines), for which we measure a width of $\simeq$19.5~\AA. Tomograms made using the outburst data, presented in \citet{pearson2006}, show that the \ion{C}{iv} emission may arise from the accretion disc, the accretion impact stream, and the irradiated donor star. Our obtained lower limit on the width of this line in quiescence is close to the value measured during outburst. This could suggest that the line is produced at a similar location. Studies of other LMXBs suggest their quiescent UV emission may arise from the heated inner edge of the accretion disc, the accretion impact stream, and the irradiated donor star \citep[e.g.,][]{froning2011,hynes2012,cackett2013_cenx4,bernardini2013}. In the following we investigate what the most likely origin of the quiescent UV emission is for EXO 0748--676.

\subsection{No disc/accretion stream in quiescence?}
Outburst studies of \exo\ in the X-ray indicated the presence of `dipping' features likely due to the presence of an ionised absorber \citep[e.g.,][]{sidoli2005broad}. \citet{psaradaki2018modelling} examined the high-resolution outburst spectra during the eclipses using the Reflection Grating Spectrometer (RGS) on board {\it XMM-Newton}. Their study found emission lines which may be due to clumps produced by the presence of the accretion stream as it impacts the disc. 
To investigate whether the \ion{C}{iv} line we detect from the source in quiescence may arise from such clumps, we extrapolated their outburst model to the FUV. We have further extrapolated this to quiescence by using the RGS data obtained during early quiescence \citep[see][for details]{diaztrigo2011}. However, no significant emission lines were present in these data and, therefore, the study carried out for the outburst spectrum could not be extended to quiescence. Furthermore, no dipping features have been observed in the X-ray during quiescence, which instead exhibited eclipses with sharp ingresses and egresses \citep{diaztrigo2011,degenaar2014probing}. This raises the question whether the obscuring structure, i.e., the disc and/or accretion stream impact point, is present in quiescence. 

Optical spectroscopy studies of \exo, carried out $\simeq$1~yr into quiescence \citep[][]{ratti2012}, showed no indications of the presence of an accretion disc (as discerned from the optical line emission). It is worth noting though, that this study was carried out early in quiescence, while our {\it HST} observations were obtained $\simeq$3.5~yr later and the disc may have built up over time. Such a build-up of the accretion disk may have been observed for another edge-on crust-cooling source, MXB~1659$-$29. For that source, this scenario was proposed based on a possible increase in the absorption column density as inferred from X-ray spectral fitting \citep[see][for details]{cackett2013change}.\footnote{This change was observed $\simeq$11~yr into quiescence and $\simeq$3~yr before MXB~1659$-$29 was seen to enter a new outburst \citep[][]{negoro2015_mxb}.} However, as argued above it is not obvious that this was the case for \exo, so we continue to explore other possibilities for the origin of is UV emission.

\subsection{Irradiation of the companion star}
Although no evidence of an accretion disc was found by \citet{ratti2012}, they observed variations at different orbital phases indicating that the irradiated face of the companion star was emitting in the optical. This raises the possibility that the quiescent UV radiation may also arise due to the irradiation of the companion by the central X-ray source. However, the observed \ion{C}{iv} emission line is much broader than the observed optical lines. We measure a lower limit of $\gtrsim 3523$~km~s$^{-1}$, whereas the reported widths of the optical spectral lines are $\simeq300-400$~km~s$^{-1}$ \citep[see table~1 of][]{ratti2012}. \footnote{We note that the \ion{C}{iv} line is actually a doublet, separated by $\simeq500$~km~s$^{-1}$, so it is possible that in our low signal to noise data two lines are merged into one. Nevertheless, the lines would still need to be significantly broader than the lines in the optical quiescent spectrum presented in \citet{ratti2012}.}

Based on their optical study in quiescence, \citet{ratti2012} suggest that \exo\ may be a black widow system, in which the pulsar wind interacts with the companion star, slowly obliterating it.\footnote{We note that the constraints on the mass of the donor star yield $>0.11~\Msun$ for a 'canonical' NS mass of $1.4~\Msun$ \citep[see][and references therein]{bassa2009}, which means that should \exo\ be an active pulsar, it would be classed as a redback, like the tMSRPs, rather than a black widow.} This scenario would require the NS in \exo\ to turn on as a radio pulsar in quiescence. It may then be similar to the handful of known transitional millisecond radio pulsars (tMSRPs) that appear to switch between radio pulsar states and low luminosity LMXB-like states in which an accretion disc is detected in the system in the optical band. 

The optical lines in the spectrum of PSR~J1023+0038, one of the tMSRPs, arise from irradiation of the companion star and are  narrow \citep[][]{mcconnell2015roche}, like the optical lines of \exo\ in quiescence \citep[][]{ratti2012}. The difference in line width makes it unlikely that the \ion{C}{iv} emission line observed for \exo\ arises due to irradiation of the companion star, but the radio pulsar scenario does pose another possibility for its origin. 

\subsection{Pulsar wind or optically thin accretion disc}
Interestingly, the UV lines of PSR J1023+0038 are also very broad ($\simeq2000-3000$~km~s$^{-1}$). Since the accretion disc in this system is truncated very far away from the NS, it was proposed that these lines are produced by the pulsar wind ionising material from the companion star \citep[i.e., the lines would then be broadened by the velocity of the wind;][]{HS2016_thesis}. Possibly a similar physical scenario can explain the broadness of the \ion{C}{iv} emission line in \exo. 

Inspecting the red wing of the airglow emission line around $\simeq$1220~\AA\ in the quiescent UV spectrum of \exo\ suggests that there may be an excess corresponding to the \ion{N}{v} emission line that was seen in outburst \citep[][]{pearson2006}. We cannot confidently constrain its presence, but if this line is there, this could lend support for the pulsar wind scenario. 
This mechanism has also been invoked to explain the UV excess in two systems hosting tMSRPs \citep{rivera2018miduv}. Nevertheless, this has only been shown for a small sample of sources and similar studies cannot be carried out for \exo\ as its continuum emission is not detected.

Another possibility, which does not require invoking \exo\ turns on as a radio pulsar in quiescence, is that material from the companion star is ionised by a geometrically thick, optically thin quiescent accretion flow. If such a flow is present in this system, we might expect material to accrete on to the NS \citep[e.g.,][]{chakrabarty2014hard,dangelo2015}. However, there are no signs of ongoing accretion in \exo.

\subsection{The (lack of) X-ray/UV correlation}
In addition to FUV spectra, we have also examined several epochs of quasi-simultaneous X-ray and NUV data in quiescence. If the UV emission arises somewhere in a quiescent accretion flow that reaches the NS, a correlation between the UV and X-ray emission would be expected. Taking into account the uncertainty of the UV spectral shape, we find no evidence for significant changes in the UV emission during quiescence, although the uncertainties are so large that changes of similar magnitude as seen in the X-rays can be hidden within the errors.

Despite that we cannot firmly determine if the X-ray and UV emission change in tandem, the X-ray/UV luminosity ratio gives additional information. We determined the ratio of quiescent X-ray to NUV emission to be $L_{\mathrm{X}}/L_{\mathrm{UV}} \gtrsim 100$. Since the UV luminosity of \exo\ is of similar magnitude as that of other LMXBs, this high ratio would suggest that it is considerably more X-ray bright than other systems (see Figure~\ref{fig:uvx}). In particular, the two NS LMXBs in the sample of \citet{hynes2012}, Cen X-4 and Aql X-1, exhibit $L_{\mathrm{X}}/L_{\mathrm{UV}} \simeq 10$. In Cen X-4, there is compelling evidence that the intrinsic NUV emission comes from the accretion flow \citep[e.g.,][]{cackett2013_cenx4,bernardini2013}, and that matter accretes onto the surface of the NS even at these low luminosities \citep[e.g.,][]{chakrabarty2014hard,dangelo2015}. The same might be true for Aql X-1, which exhibits X-ray variability of several magnitudes in quiescence \citep[e.g.,][]{rutledge2001,campana2003_aqlx1,cackett2011_aqlx1,cotizelati2014}. The fact that \exo\ has a similar UV luminosity as these systems but is much brighter in the X-rays, would then suggest that its X-ray emission is not (predominantly) powered by low-level accretion.  \\
\\

\noindent In conclusion, our study of (quasi-)simultaneous UV and X-ray observations of \exo\ in quiescence hints that the observed broad UV emission line arises from ionisation of material from the companion star, possibly by a pulsar wind. Similar to previous optical and X-ray studies, our combined UV/X-ray analysis does not reveal evidence for the presence of a quiescent accretion stream, nor that the high quiescent X-ray emission of \exo\ arises from ongoing low-level accretion. Our results thus seem to be in favor of explaining the bulk of the quiescent X-ray emission of \exo\ as cooling of the NS crust. This implies that such studies can be used to infer the properties of the crust and core of this NS \citep[e.g.,][]{degenaar2011further,parikh2020dip}.

\section*{Acknowledgements}
The authors are very grateful to the referee, Craig Heinke, for a valuable report that helped improve this manuscript. AP, ND and JVHS are supported by a Vidi grant awarded to ND by the Netherlands Organization for Scientific Research (NWO). IP and EC are supported by the Vidi grant 639.042.525. JVHS  acknowledges funds from a Science and Technology Facilities Council grant ST/R000824/1 research fellowship. Further support for \hst\ program GO-13108 was provided by NASA through a grant from the STScI. 

\section*{Data Availability}
The data underlying this article are available in Zenodo, at https://doi.org/10.5281/zenodo.3908291. The datasets were derived from sources in the public domain: \url{https://archive.stsci.edu/hst/search.php}, \url{http://nxsa.esac.esa.int/nxsa-web/\#home}, and \url{https://heasarc.gsfc.nasa.gov/cgi-bin/W3Browse/swift.pl}.




\bibliographystyle{mnras}



\appendix

\section{UV spectral shape corrections}\label{sec:A1}

\begin{figure*}
 \begin{center}     
  \includegraphics[width=\columnwidth]{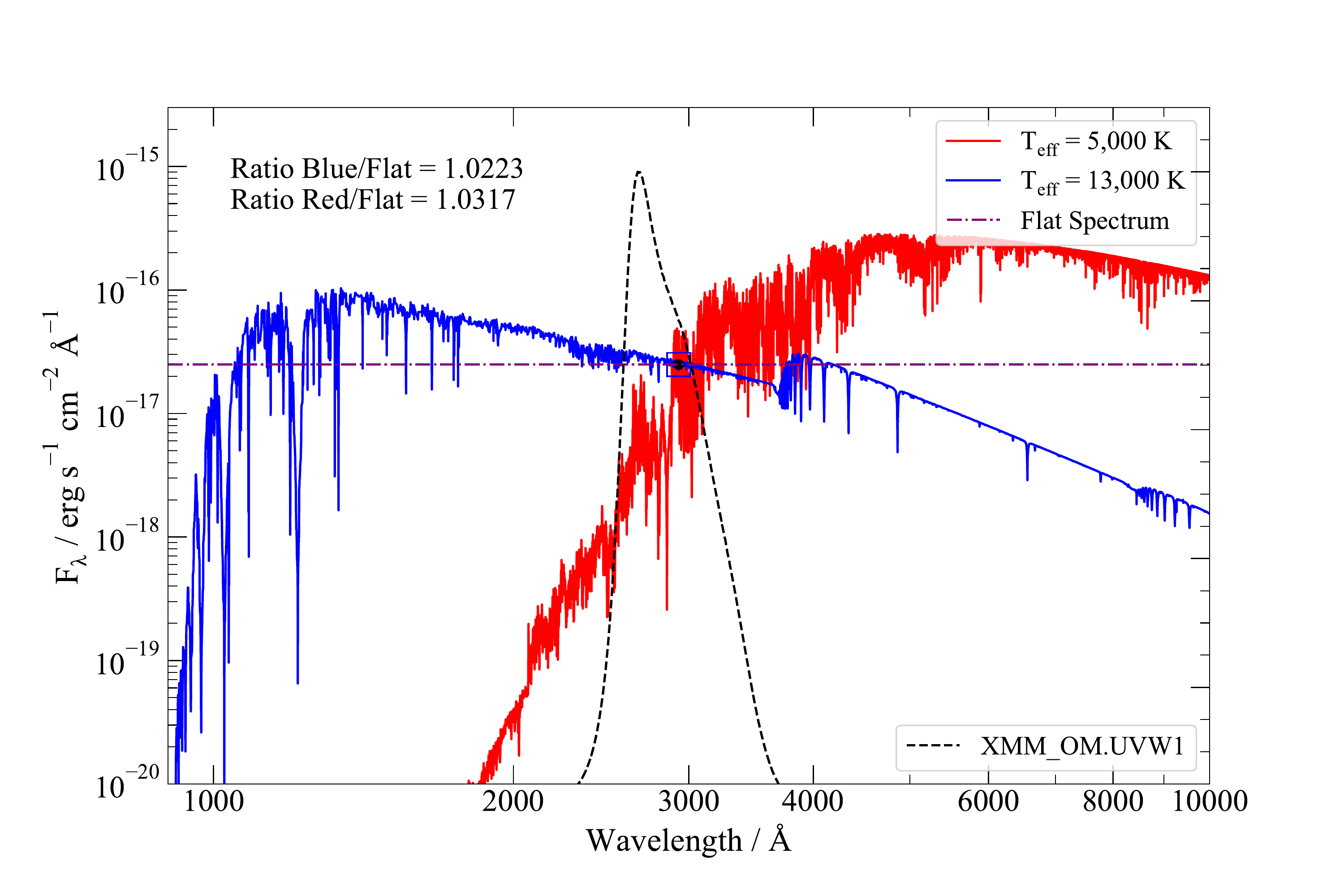}
    \includegraphics[width=\columnwidth]{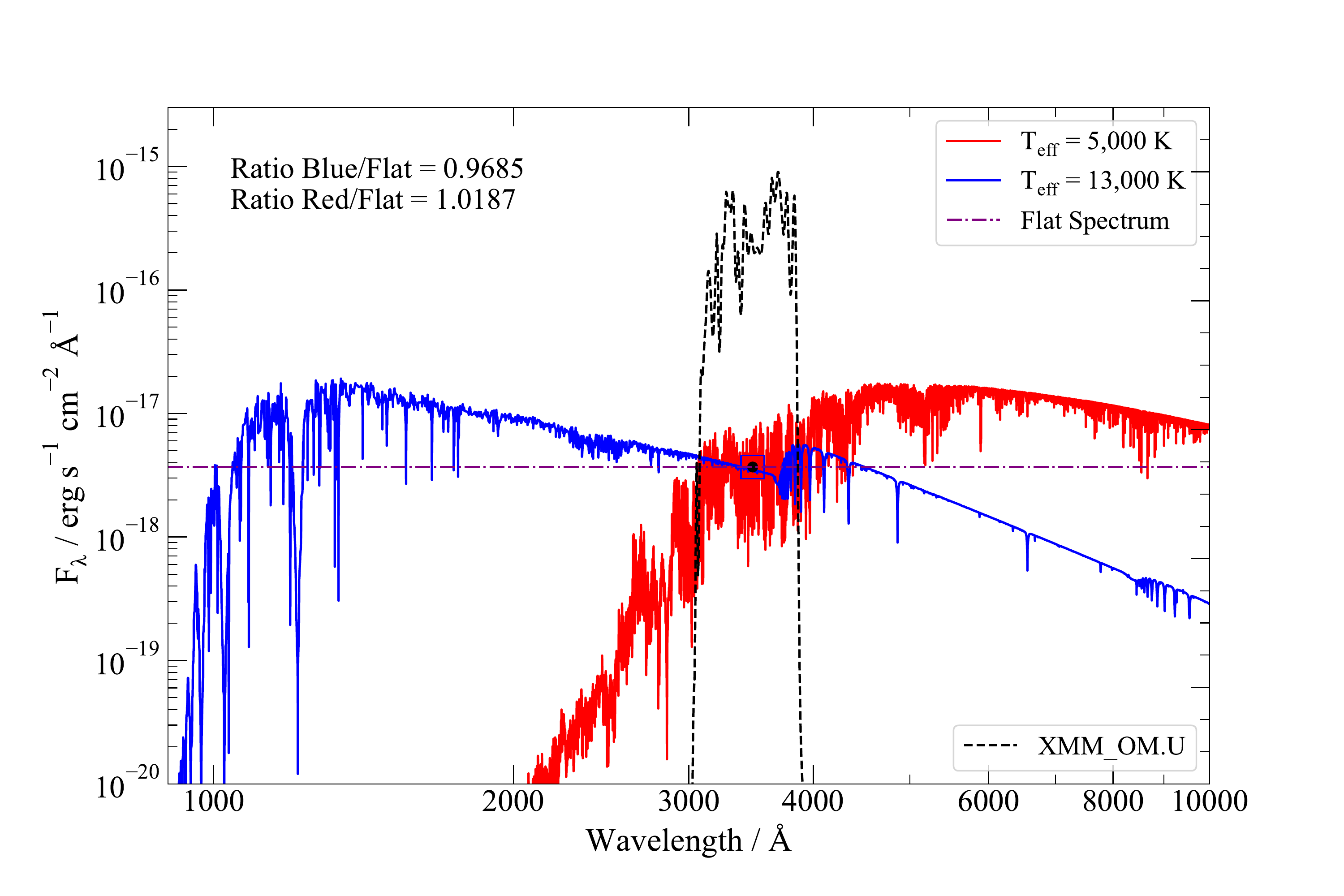}
        \includegraphics[width=\columnwidth]{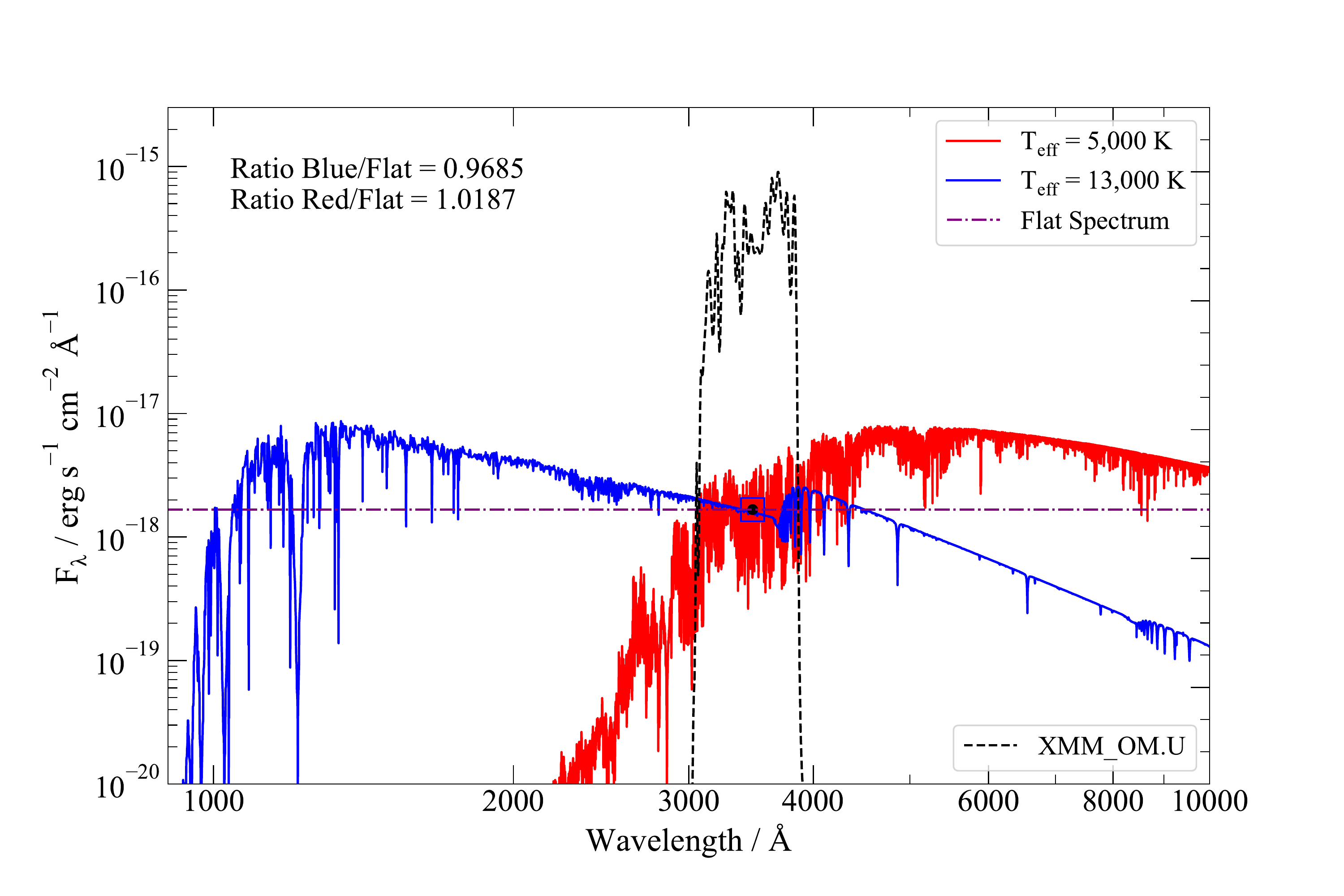}
            \includegraphics[width=\columnwidth]{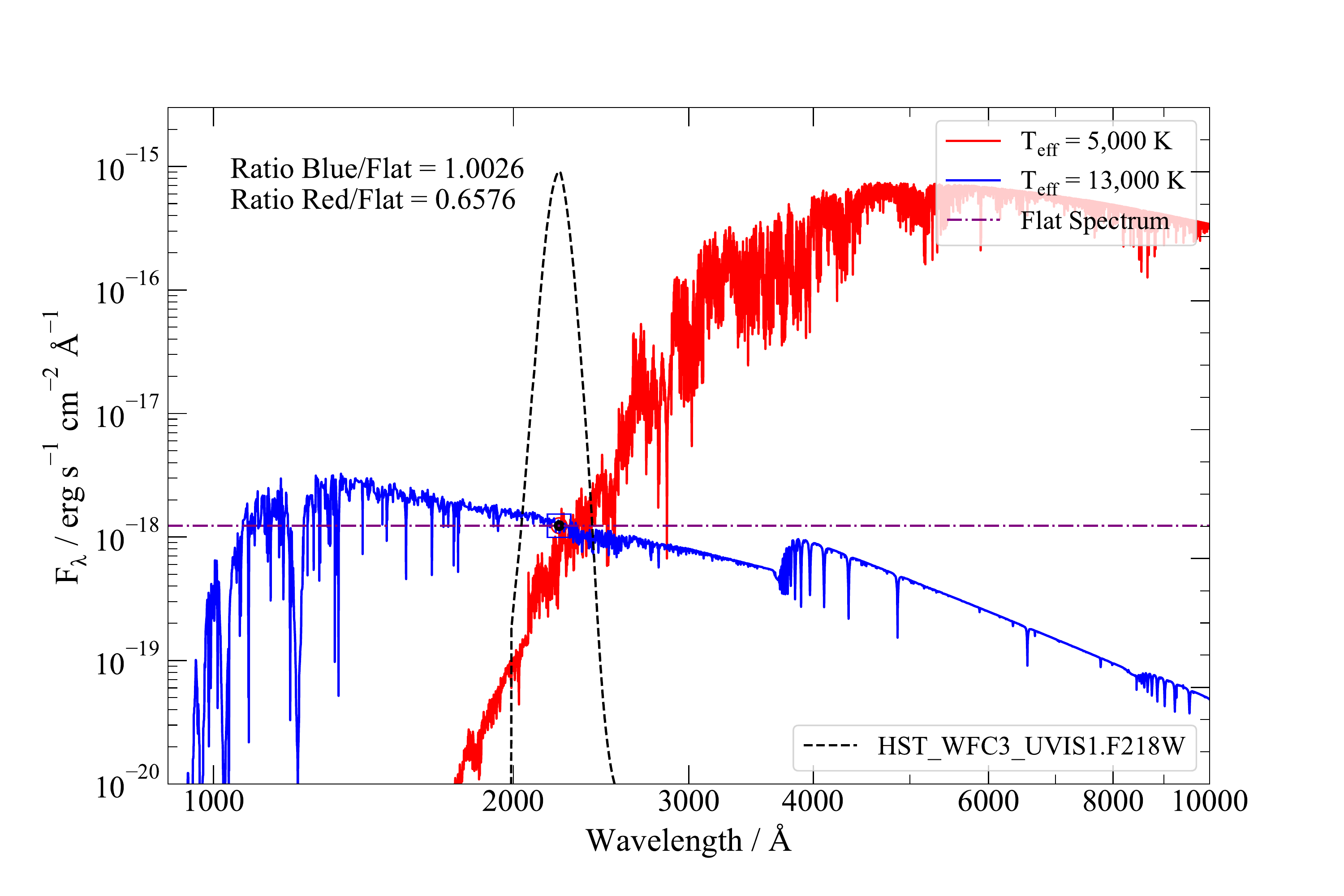}
                \includegraphics[width=\columnwidth]{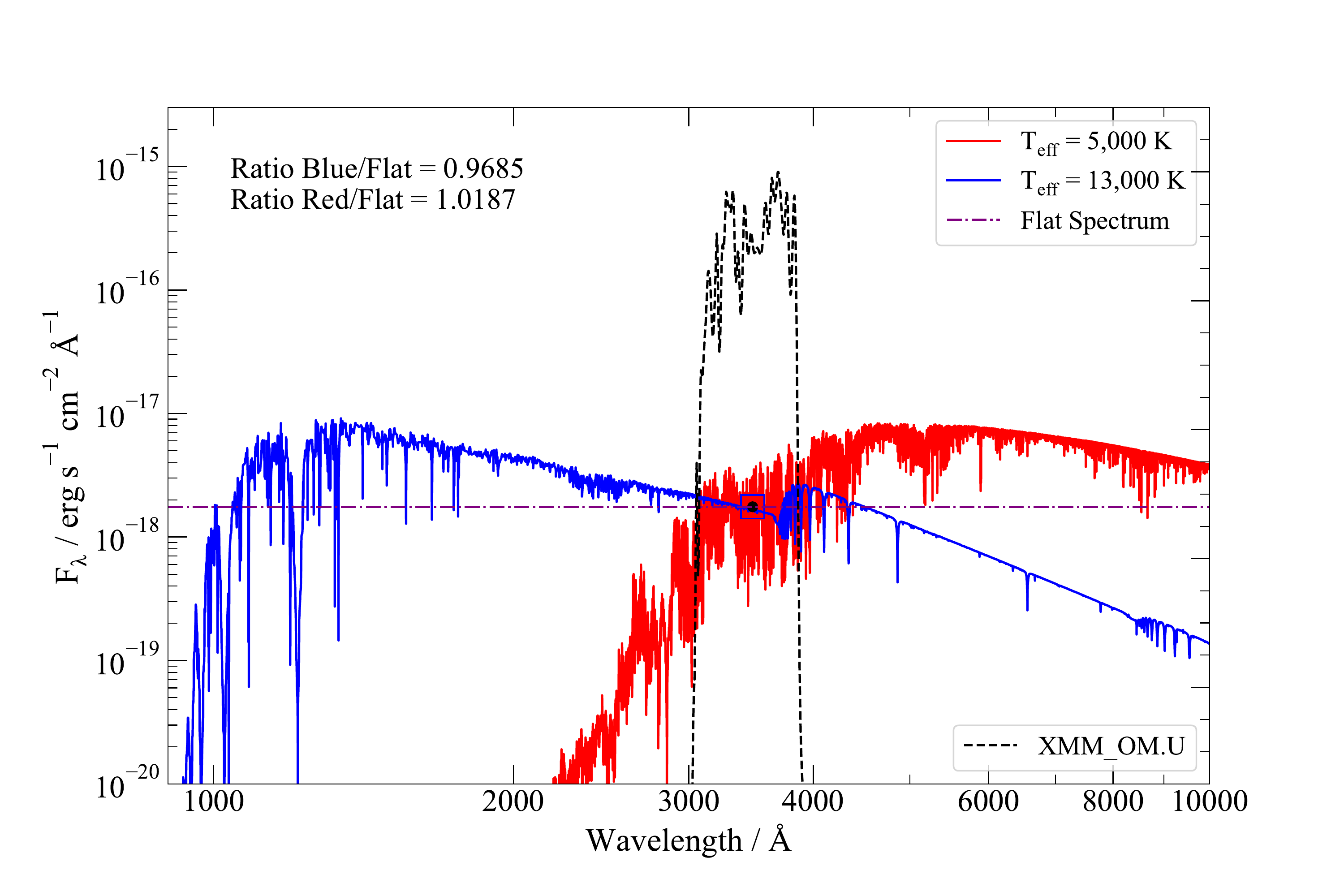}
                    \includegraphics[width=\columnwidth]{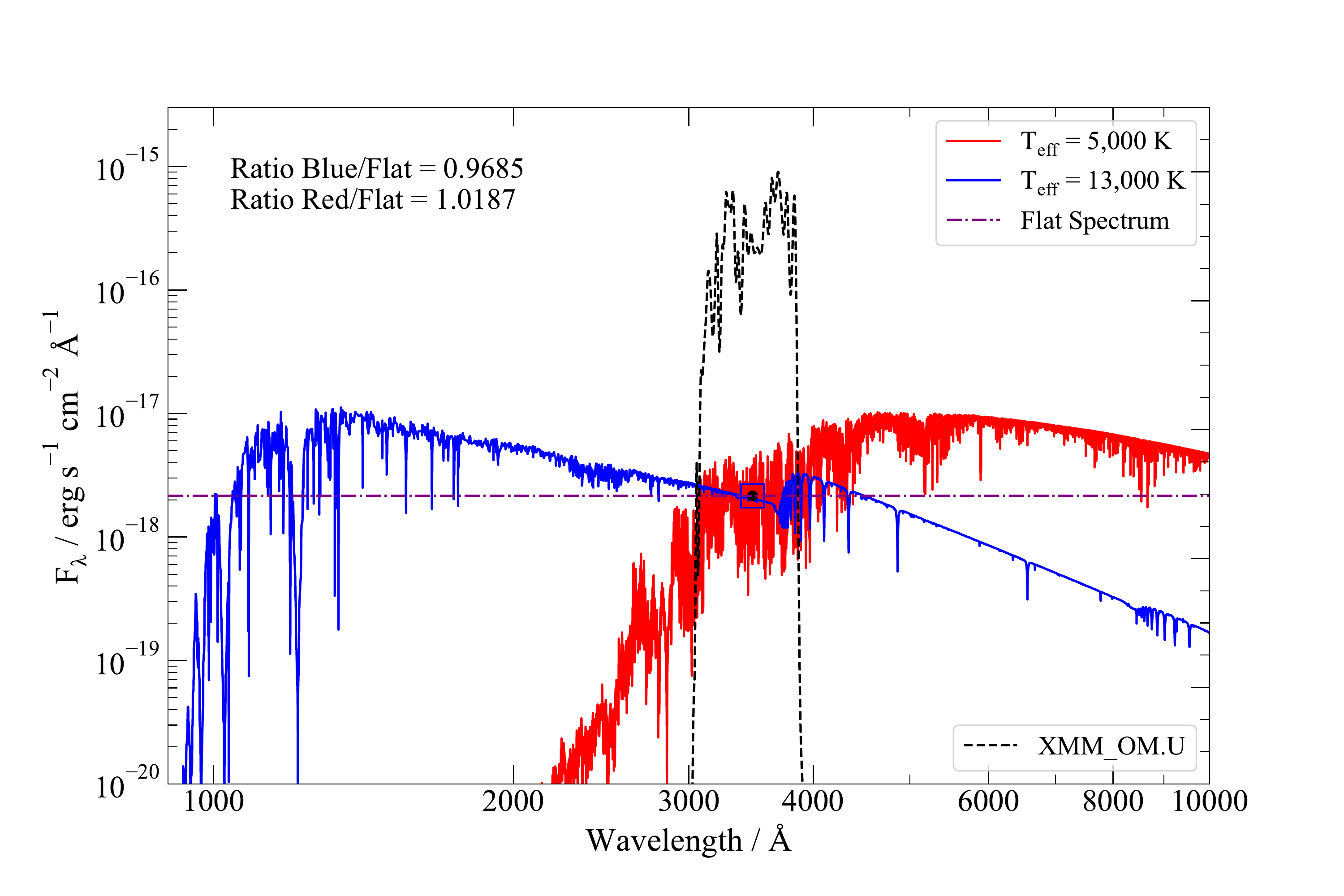}
    \end{center}
\caption{ Estimates of the effect of the unknown UV spectral shape on the inferred UV luminosity. For each observation we show the measured UV flux (upper limit) as the black data point, while the blue and red curves represent stellar-atmosphere spectra with effective temperatures of 13\,000~K and 5\,000~K, respectively. The black dashed curves show the different filter transmissions. Printed in the top left are the ratios of the integrated fluxes for these models compared to a flat distribution, indicated by the dash-dotted horizontal purple lines, over the filter passband. These plots show that the unknown spectral shape has a relatively minor impact on the calculated U-band luminosity (of a few per cent), but has a larger impact for bluer filters.}
 \label{fig:spec_effect}
\end{figure*}


\bsp	
\label{lastpage}
\end{document}